\documentclass[12pt,preprint]{aastex}

\shorttitle{MHD Study of CME-Comet Interaction} \shortauthors{Jia et
al.}

\begin{document}


\title{Study of the April 20, 2007 CME-Comet Interaction Event \\ with an MHD Model}


\author{Y. D. Jia\altaffilmark{1}, C. T. Russell\altaffilmark{1}, L. K.
Jian\altaffilmark{1}, W. B. Manchester\altaffilmark{2}, O.
Cohen\altaffilmark{2}, A. Vourlidas\altaffilmark{3}, K. C.
Hansen\altaffilmark{2}, M. R. Combi\altaffilmark{2}, T. I.
Gombosi\altaffilmark{2}}

\altaffiltext{1}{IGPP, University of California, Los Angeles, CA
90095} \altaffiltext{2}{CSEM, University of Michigan, Ann Arbor, MI,
48109} \altaffiltext{3}{Solar Physics Branch, Space Science
Division, Naval Research Laboratory, Washington, DC 20375}


\begin{abstract}
This study examines the tail disconnection event on April 20, 2007
on comet 2P/Encke, caused by a coronal mass ejection (CME) at a
heliocentric distance of 0.34 AU. During their interaction, both the
CME and the comet are visible with high temporal and spatial
resolution by the STEREO-A spacecraft. Previously, only current
sheets or shocks have been accepted as possible reasons for comet
tail disconnections, so it is puzzling that the CME caused this
event. The MHD simulation presented in this work reproduces the
interaction process and demonstrates how the CME triggered a tail
disconnection in the April 20 event. It is found that the CME
disturbs the comet with a combination of a $180^\circ$ sudden
rotation of the interplanetary magnetic field (IMF), followed by a
$90^\circ$ gradual rotation. Such an interpretation applies our
understanding of solar wind-comet interactions to determine the
\textit{in situ} IMF orientation of the CME encountering Encke.
\end{abstract}



\section{Introduction}

The atmosphere of a comet continuously interacts with the solar
wind, which in turn varies continually in the frame of the comet.
This variability makes both real and apparent changes in the
appearance of cometary plasma tails. Better knowledge of such
interaction processes enhances our ability to infer the \emph{in
situ} solar wind conditions from their observed behavior.

Disconnection events (DEs) are the most dramatic interactions
between the solar wind and comets. During such events, the entire
plasma tail is uprooted while a new tail grows, usually accompanied
by the appearance and disappearance of tail rays. As summarized by
\cite{Voelzke05}, two types of solar wind structures are believed to
be responsible for the onset of DEs: interplanetary shocks and
heliospheric current sheets (HCS).

In recent years, a third source of tail disconnections has been
observed. Coronal mass ejections (CMEs) also appear to be able to
cause DEs, but how they create DEs has never been fully understood
(e.g. \cite{Buffington08,Kuchar08}). On April 20, 2007, Heliospheric
Imager-1 (HI-1) of the Sun-Earth Connection Coronal and Heliospheric
Investigation (SECCHI) aboard the STEREO-A spacecraft
\cite[]{Howard08} captured a DE when a CME hit comet 2P/Encke at
0.34AU. Both the traveling CME and the variation in the Encke plasma
tail are visible with high spatial and temporal resolution,
providing an opportunity to study the nature of such processes
\cite[]{Vourlidas07}. This event is hereafter called "the April 20
event".

In this letter, the interaction sequence is examined and reproduced
by magnetohydrodynamic (MHD) modeling. Our result supports that the
tail disconnection is caused by the flux rope structure (e.g.
\cite[]{Russell08}) embedded in the CME. The real-time interaction
of a flux rope with comet Encke in our simulation explains both the
unusual appearance and time evolution recorded by STEREO. Herein we
attempt to reconstruct the CME from both HI-1 images and the
cometary response, which together constrain the \textit{in situ}
magnetic field orientation in the CME at 0.34 AU.

\section{Observations}

Comet 2P/Encke is a Jupiter family comet with an orbital period of
3.3 years and a perihelion production rate of approximately
$2\sim6\times10^{28}\,s^{-1}$ \cite[]{Makinen01,Lisse05}. On April
20, 2007, comet Encke was traveling from north to south at an
orbital speed of approximately $60\,km/s$, having just passed
perihelion close to the ecliptic plane (See
http://ssd.jpl.nasa.gov/horizons.cgi). During the April 20 event,
the STEREO-A spacecraft was in its heliocentric orbit close to the
Earth. The HI-1 measures Thomson scattered light with a passband of
$630\sim730\,nm$.

Figure 1 shows the interaction stages with images extracted from the
online animation in \cite[]{Vourlidas07}. The sun is to the left,
while the solar wind flows to the right. The tail of comet Encke is
the bright line in the center, with the head lying at the left. In
each image, there is a vertical bright front and its succeeding dark
region. The bright cloud represents the sheath while the dark region
is a possible flux rope. Also note that the less bright region at
the left side of each images is not the end of the CME. At the time
of the interaction, the CME diameter is approximately 0.14 AU along
the equator and 0.21 AU along the north-south direction and its
center is located at 48 $R_s$ from the Sun. Note that these are
projected quantities along the sky plane and are fairly typical for
such events (e.g., \citep{Vourli00, Subra07}). If we assume
self-similar expansion, we estimate that the structure was about 0.6
$R_s$ in diameter when it erupted. The larger size along the normal
to the equator implies that the CME is 'pancaking' as has been
predicted by several models (e.g., \citep{Riley04}). These
measurements relate to the outer envelope of the CME which includes
overlying streamer material and likely overestimates the actual size
of the magnetic fluxrope. Since this is generally considered to be
the dark void in white light CMEs, we estimate a fluxrope size of
0.06 AU by measuring the dark region behind the bright front. This
diameter is consistent with the diameter estimated from the time
elapsed for the dark region to pass the comet with a speed of
$500km/s$.

The diameter, as estimated above, is the lower limit of this flux
rope but is significantly smaller than the average diameter at this
heliocentric distance (e.g., \cite[]{Lepping07,Jian08}). However,
the size of the CME is consistent with similar measurements in the
LASCO coronagraphs. Besides, our calculations are based on direct
measurements of CME features while in-situ estimates are based on
model fits to the magnetic data and could suffer significant
uncertainties. On the other hand, numerous small scale fluxropes
have been detected with the STEREO in-situ instruments
\citep{Kilpua09}. Therefore, our estimates for the fluxrope size are
reasonable and can use them in the simulations with confidence.

At 18:10 UT, the leading side of the CME flux rope is
1$\times10^6km$ downstream of the comet head. At 18:50 UT, the comet
tail is elongated while starting to break away from the head. At
19:30 UT, the comet head is engulfed in the flux rope region of the
CME, while the tail is completely disconnected. At 20:10 UT, the gap
between the disconnected tail and the comet head equals over three
million kilometers.

Both the slow solar wind and the interplanetary magnetic field (IMF)
were in the radial direction. The projected CME speed is about
$500\,km/s$. The dynamic pressure changed less than 20\% due to the
CME. Although recognizable in the images, the density variation in
the CME is $<$1\%, so this DE was not caused by pressure effects
\cite{Vourlidas07}. This event is puzzling for the following
reasons:

First, this event appears to be different from past events. There is
a long gap developed between the old tail and the end of the new
tail. In addition, this DE is accompanied with tail elongation with
no tail rays.

Second, this DE must have been caused by a mechanism other than a
HCS crossing or shock impact, because there were neither HCS
crossings nor shocks seen in the HI images. In this study we discuss
how the flux rope of the CME could have caused the tail
disconnection.

In the top panel of Figure 1, the CME front has caused a kink in the
comet tail three million kilometers from its head. Although this
kink is not modeled in our study, it does suggest that the
disconnection is not caused by the CME front. The comet tail is
aligned in the solar wind flow direction, so the CME encounters the
comet head before it reaches the tail. At 18:50 UT, the DE becomes
visible while the comet head has been in the flux rope for an hour.
This scenario is consistent with the appearance of a DE caused by a
$180^\circ$ rotation in the IMF, with the observing line of sight
perpendicular to the IMF plane \citep[panel b, Figure 4]{Jia07}. At
19:30 UT and 20:10 UT in Figure \ref{fig1}, the new tail appears
consistent with the appearance of shortened comet tails when the
observer looks through the thin direction of the current sheet
\citep{Russell91GMS}.
\section{MHD Model}

In this study we use the Block Adaptive Tree Solar-wind Roe Upwind
Scheme (BATS-R-US), a 3-D global MHD numerical code
\cite[]{Powell99}, and the Space Weather Modeling Framework (SWMF)
that employs BATS-R-US for several of the coupled physical domain
modules \cite[]{Toth05}.

Here, the solar corona and inner heliosphere models (based on
BATS-R-US) are run with the SWMF code to predict the ambient solar
wind condition at 0.34 AU \cite[]{Cohen07}. The estimated IMF
magnitude is 20 nT, the density is 20 $amu/cm^3$, the velocity is
440 $km/s$ and the temperature is $3\times10^5\,K$. Such a solar
wind condition is adopted as the outer boundary for the comet code.

A previous version of the BATS-R-US comet code was used for a
time-dependent generalized DE study with HCS crossings
\citep{Jia07}. In this study the same code is applied to simulate
the time evolution as shown in next section. The x-axis points along
the solar wind flow direction, z-axis points north, and the y-axis
completes the right-hand system. The calculation domain is
$2.4\times0.4\times0.4$ in unit of $10^6\,km$, with finest
resolution of 25 km in the cometary contact surface region, and a
resolution of 800 km in the cometary tail.

For simplicity, both the radial component of the background IMF and
the temperature variations in the CME are neglected. The CME
structure is represented by a simplified flux rope that is
5.4$\times10^6\,km$ across. This value is smaller than the estimated
size of 9$\times10^6\,km$, to qualitatively simulate the shorter
path that the comet travels in the CME, because in this event the
comet interacts with the side or even the leg of the fluxrope
\cite{Vourlidas07}. The axis of the flux rope is in the y direction.
The density and velocity variations in the CME are neglected. In
addition, we assume that comet Encke crossed the axis of the CME.
The dimensions of the bow wave of comet Encke are estimated to be
less than 0.2 $\times10^6\,km$ (the subsolar distance is
$5\times10^4\,km$), which is orders of magnitude smaller than the
diameter of the flux rope, of which the curvature of field lines is
negligible. Consequently, the interaction process is modeled simply
as the effect of the rotation of the IMF vector. At initial state,
the function of the magnetic field is:
\begin{eqnarray}
\label{eq:FR} & & B_x = 0 \,nT \nonumber \\
&&   B_y = \left\{
      \begin{array}{ll}
         0 & +\infty \geqslant x \geqslant -0.85 \\
         20sin\left( {{{\pi\over2}\left(x+0.85\right)}\over{\left(-3.55+0.85\right)}} \right) & -0.85 > x\, >
         \,-3.55 \\
         20 & else
      \end{array} \right. \nonumber \\
& &   B_z = \left\{
      \begin{array}{ll}
         20 & +\infty \geqslant x \geqslant -0.4 \\
         -20 & -0.4 > x \geqslant -0.85 \\
         20cos\left( {{{\pi\over2}\left(x+0.85\right)}\over{-3.55+0.85}} \right) & -0.85 > x\, >
         \,-3.55 \\
         0 & else
      \end{array} \right. \nonumber,
\end{eqnarray}
where x is the x-coordinate in unit of $10^6km$, $B$ is in unit of
$nT$. The parameters, 0.4 indicates the upstream boundary of our
simulation box at $-0.4\times10^6km$. At 15:45 UT, the front current
sheet is located at $-0.85\times10^6km$, with a thickness of
$0.35\times10^6km$. Thus the center of the flux rope with a radius
of $2.7\times10^6km$ is at $-3.55\times10^6km$. These parameters are
selected to reproduce the general dynamics, while for a future work
that compares the timing of this evolution, a more accurate set of
parameters need to be tested.

\section{Model Results}

The model calculation of the interaction process between this CME
and the comet is shown in 2-D projections of Figure \ref{fig3}. The
color contours of plasma density show the location of the bow shock
and tail. The black lines represent magnetic field lines. Left
panels show the meridional plane, while right panels show the
ecliptic plane. Please note that these volume densities shown in 2-D
slices are not yet directly comparable to the integrated column
densities in the HI images.

At 16:10 UT, the front of the flux rope (marked by the dashed
vertical white lines) is at $-10^5\, km$, in front of the comet bow
shock. The IMF is in the x-z plane so the tail appears thinner in
this plane, and wider in the x-y plane. The comet appears similar to
its initial state.

At 16:30 UT, marked by the white dashed lines, the interface of
reversed field lines has formed a cone centering at the comet head.
In the x-y plane, the axial component in the flux rope starts to
increase behind the front, as marked by the black field lines to the
left of the white dashed line. The field to the right of the white
line is the old IMF pointing northward. The comet tail appears
stretched in both slices. After this, the white dashed lines in
panels a and b continue to fold toward the tail as this stretched
interface propagates downstream.

At 17:40 UT, the tip of the interface between the reversed field
lines has propagated to 400,000km, while the axis of the flux rope
has entered the upstream boundary of the simulation domain (not
shown). Magnetic reconnection in the tail has generated a density
pileup region, which appear as the trailing head of the disconnected
tail, as studied by \cite{Jia07}. At this moment the pile up region
has propagated to 300,000 km, as marked by the density increase
surrounded by the field line on the right. To the left of the
density pile-up, a new tail starts to grow. This new tail becomes
wider in the x-z plane and thinner in the x-y plane, as a
consequence of field rotation in the IMF. The field to the left of
the white lines has both y and z components. The old field lines
pointing northward are confined in the tail to the right of the
white lines. Better interpretation of the detailed evolution
requires higher resolution and resistive MHD, or Hall MHD that
treats the reconnection in a more realistic way.

At 18:30 UT, the axis of the flux rope is significantly bent, as
marked by the red dotted line in the left panel. At the top
boundary, the axis of the flux rope has passed through the shown
region. Along the x-axis, the center of the flux rope has only
propagated to the inner coma, while the front of the flux rope has
also passed through the shown region. The IMF along this red line is
primarily in the $y$ direction (zxis component), while those to the
right of the red line are rotating from the $y$ direction to the $z$
direction as they leave the red line. The red line in the right
panel marks the interface of the field lines with the same
orientation. This stretched red line suggests that the highly draped
field lines significantly affects the force balance in the
interaction process, and thus controls the evolution speed of this
process. In the next step for a study to match the observation in
real-time, the actual structure in the CME field need to be
considered.

The density pile-up is at 500,000 km, as marked by the right-most
field lines. To the left of the density pile-up, the developing new
tail appears wider in the x-z plane and thinner in the x-y plane,
consistent with the result of a $90^\circ$ field rotation.

\begin{figure}
\epsscale{.60}
\plotone{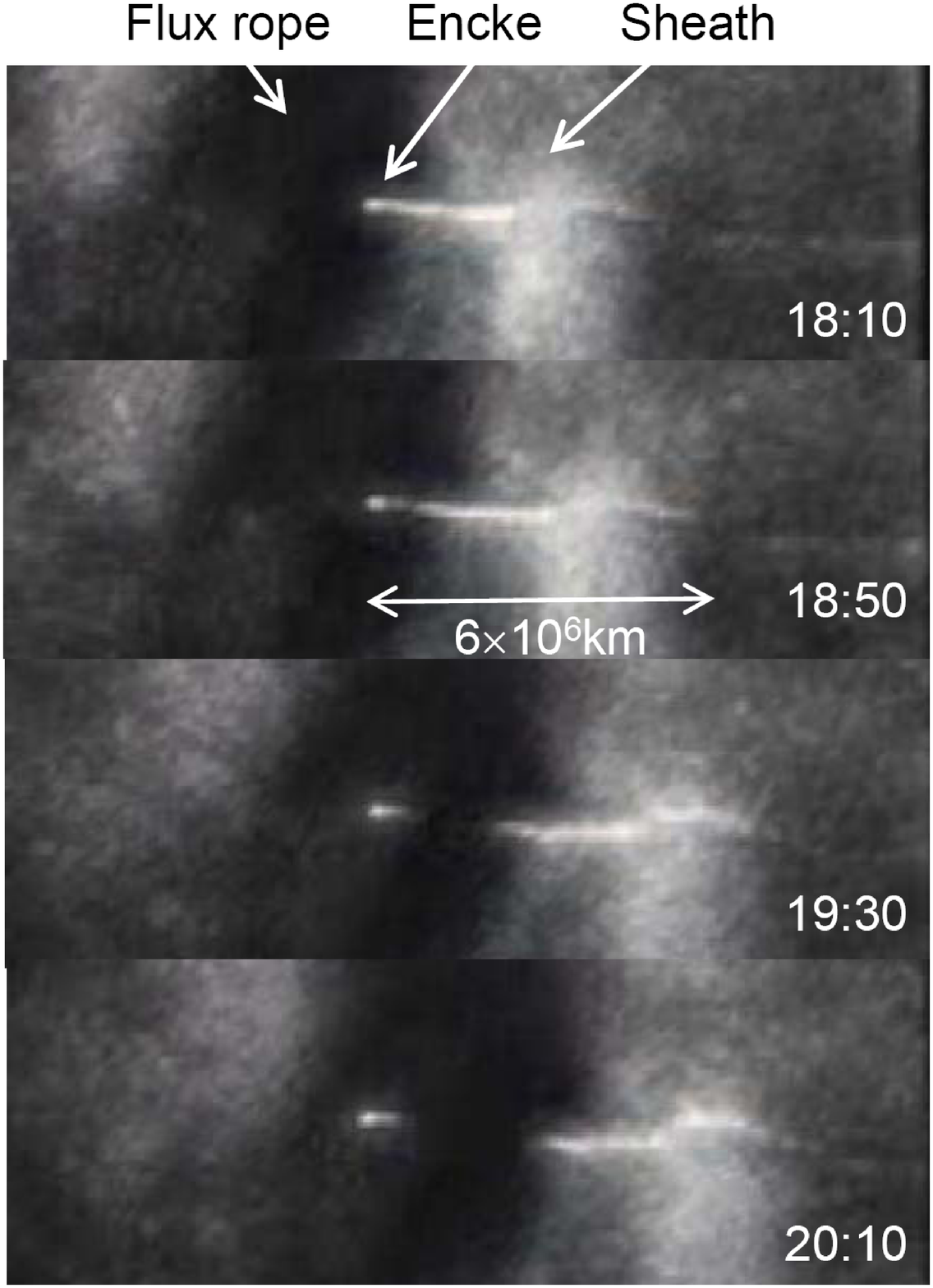} \caption{Time sequence of the CME-comet interaction
process observed by the STEREO A HI-1. Brightness represents
relative density. Solar wind comes from the left, while comet Encke
and its tail is at the center pointing to the right. As marked by
the arrows, the bright cloud shows the front sheath of the CME,
while the dark region shows the flux rope. Images extracted and
enhanced from online material by \cite{Vourlidas07}. \label{fig1}}
\end{figure}

\begin{figure}
\includegraphics[width=20pc]{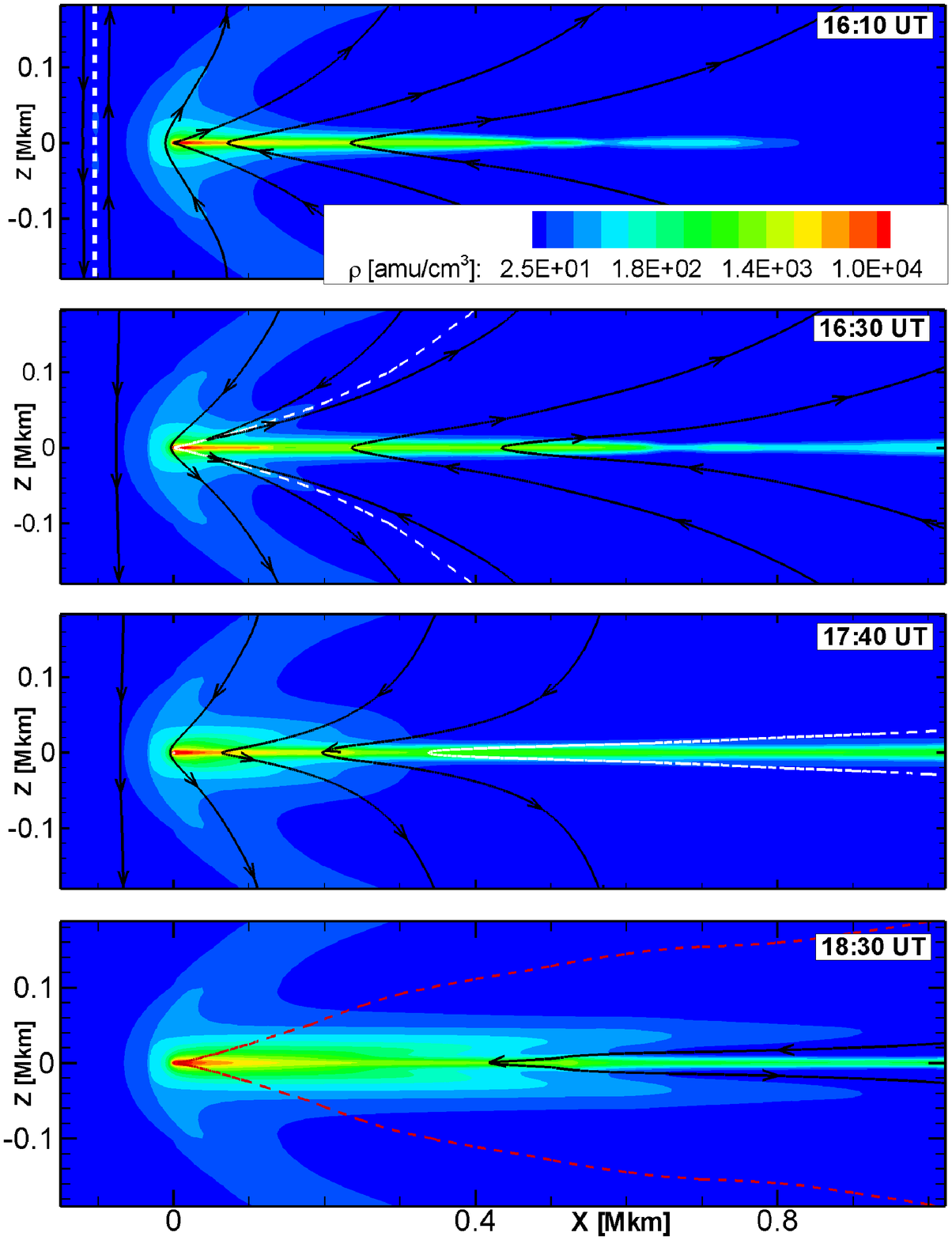}\includegraphics[width=20pc]{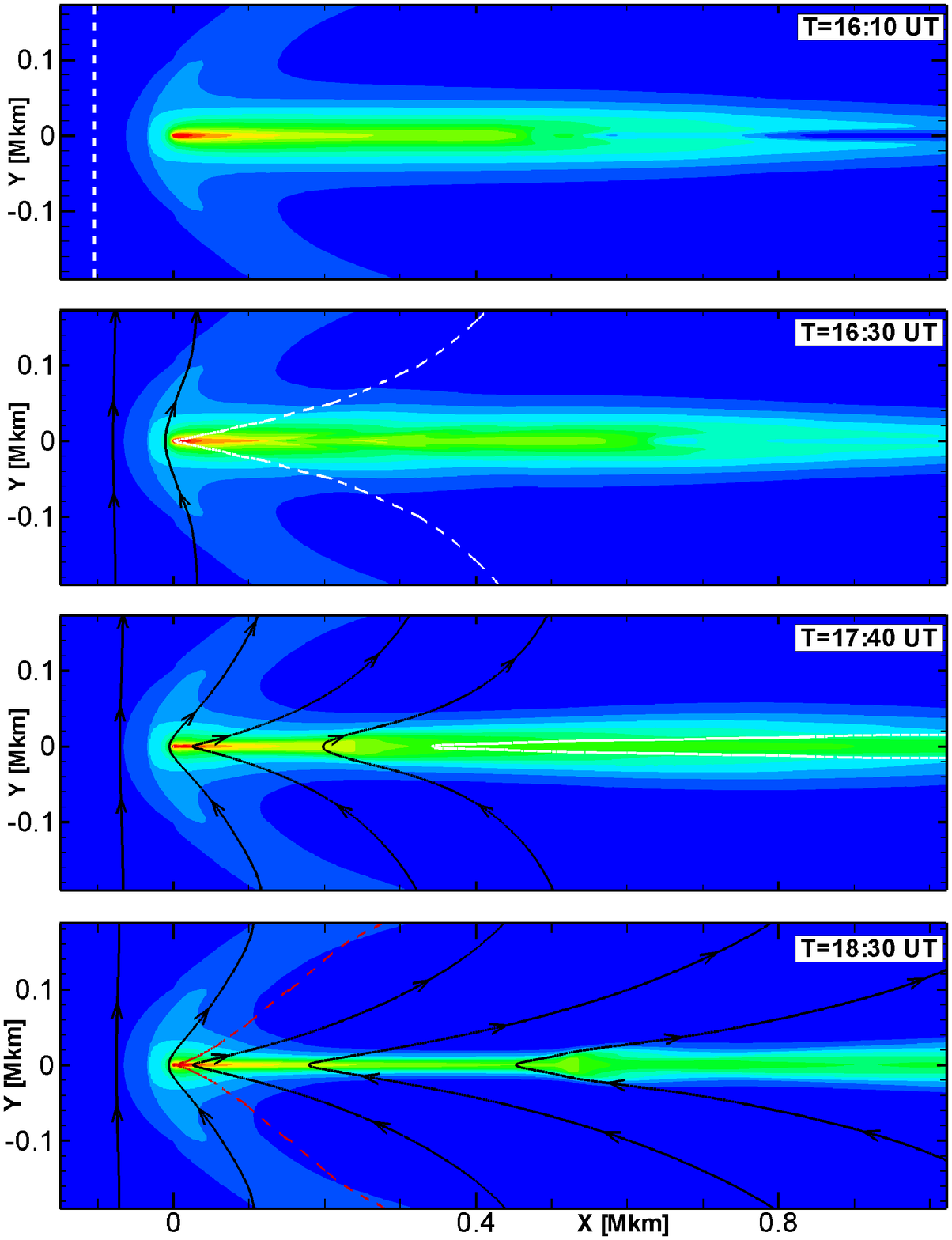}
\caption{The time evolution of the
disconnection event in 2-D slices. Color contours represent plasma
density showing the bow shock and tail. The black lines represent
magnetic field lines. White dotted lines show the current sheets,
while red dashed lines show the location of the flux rope axis. Left
panels show the meridional plane, while panels on the right show the
ecliptic plane. \label{fig3}}
\end{figure}

\begin{figure}
\includegraphics[angle=-90,width=39pc]{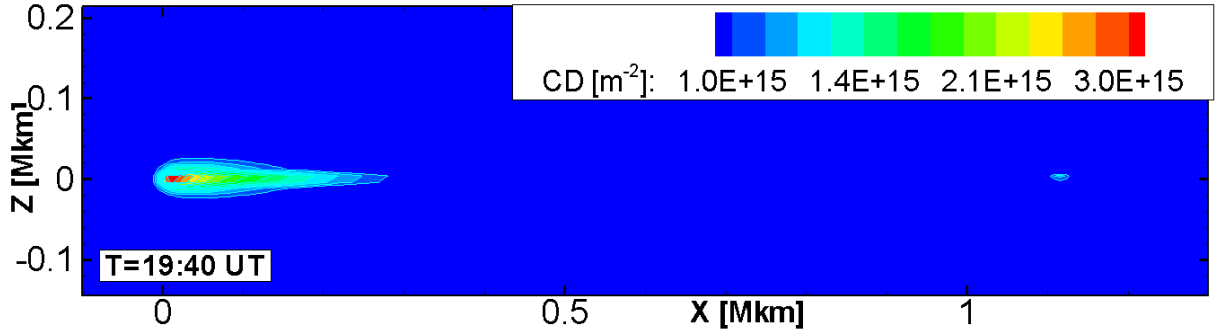} \caption{The column density
along the y-axis at 2:30h. This figure is a still image from an MPEG
animation of the interaction process available in the electronic
edition. \label{fig4}}
\end{figure}

The column-integrated density contours measured along the
y-direction are shown in Figure \ref{fig4}. The head of a
disconnected tail has propagated to x=1.1$\times10^6$ kilometers,
while a short new tail is formed and stays short. The complete
evolution process is available as an MPEG animation in the
electronic edition of the {\it Astrophysical Journal}. Between 16:00
UT and 17:20 UT, the length of the tail keeps growing, until the far
tail becomes too weak to be visible. These earlier phases resembles
the top image in Figure \ref{fig1}, where the comet tail is
stretched but not separated. At 16:40 UT, the density pile-up starts
to form at 50,000 km. At 17:00 UT, this region appears separated
from the head and propagated to 70,000 km. The separation is only
visible from the decrease in column density contours, but not
directly observable yet, because it still overlaps with the new
tail. At 17:20 UT, the pile-up is at 120,000 km, while the tail is
stretched to its maximum length. From 17:40 UT to 18:00 UT, the
density increase is at 200,000 km, while the brightness of the new
tail close to the density increase is still higher than the old
tail.

At 18:20 UT, the new tail appears short while the density increase
is at 500,000km, separated from the old tail. After 18:40 UT, the
new tail stays the same length, which is shorter than the one at
18:20 UT. The density increase, as the trailing head of the
disconnected tail, propagates from 600,000 km to $1.1\times10^6\,km$
from 18:40 UT to 19:40 UT, resembling the phase between 18:10 UT and
19:30 UT, as shown in the two middle images of Figure 1. Afterward,
the tail remains short while the old tail recesses, resembling the
two lower panels in Figure \ref{fig1}. The simulation can be
improved with a larger calculation domain and higher resolution in
the tail to simulate the complete process, with an extension of
$10\times10^6\,km$, and a two-species code to track the heavy ions
to achieve a higher contrast ratio to better observe the tails. Even
with its present limitations, the comparison between this modeled
time sequence and the STEREO images show sufficiently similar
evolution to support our interpretation.

For comparison, both a test case with only a $180^\circ$ IMF sudden
rotation and a case with $90^\circ$ gradual rotation in 1.5 hours
were studied. The $180^\circ$ case produces a DE with a similar
evolution speed, although not visible along the y-axis. The
$90^\circ$ case does not produce a DE. These two cases is comparable
with case 1 and case 2 for a Halley-sized comet (35 times more gas
production and thus evolves slower) presented by \cite{Jia07}.

The production rate of comet Encke used in the simulation above is
the lower limit of its perihelion value during its past 4 perihelion
passes. A case with its upper limit, $6\times10^{28}$ was also
studied (not shown here). In the upper limit case, the evolution
exhibited similar processes, but with a stronger brightness and
slower pace. Development of Hall MHD or resistive MHD models for
such interactions, plus the analysis of more events, is expected to
better reproduce the reconnection rate and to better address the
evolution speed.

\section{Discussion}

In summary, the MHD simulation produces magnetic reconnection in the
comet tail and provides an answer to how the CME triggered a DE in
the April 20 event. The comet tail evolution results from a
combination of a $180^\circ$ IMF sudden rotation that causes
front-side and then tail-side magnetic reconnection, with a
$90^\circ$ gradual rotation that reduces the number of cometary ions
along the line of sight. Such an interpretation applies our
understanding of solar wind-comet interactions to better constrain
the \textit{in situ} IMF orientation of this CME at 0.34 AU.

Compared with the HI-1 observation, there are still differences in
detail from our idealized simulation, suggesting our numerical model
should be improved. In particular, instead of a single species comet
model, a two-species model that tracks both the solar wind ions and
the cometary heavy ions is needed to better reveal the column
density of cometary ions.

The modeled evolution pace of this DE is slower than observed. In
addition to the limitation of ideal MHD, there are three possible
reasons that lead to this difference. First, the comet may not pass
the axis of the CME, indicating an uncertainty in the length of the
path in the flux rope than assumed. In addition, the comet passed
through the flank of the flux rope rather than the axis
\cite[]{Vourlidas07}. The asymmetry in the flux rope may affect the
magnetic pressure during the evolution. Second, the real 3-D field
and plasma conditions of the flux rope may provide stronger
acceleration force to the disconnected tail. In this case, the
curvature of azimuthal field may cause stronger tension force while
interacting with the cometary structures. Lastly, a stronger IMF, or
more complex field and plasma fluctuations in the flux rope may
speed up this process.

The IMF conditions behind the axis of the flux rope are not examined
in this model. From this disconnection event, the cometary tail
grows longer after the CME has passed, indicating that the IMF is
back in the x-z direction.

It should be noted that this study is limited to one type of
CME-comet interaction mechanism. Clearly, CME-comet interactions
involve several different physical processes, and the observed DEs
are formed in a variety of ways. We need to gather more \textit{in
situ} data at the times of cometary tail disconnections to
understand fully the range of interaction leading to these
disconnection events.


\bibliographystyle{apj} 

\clearpage

\end{document}